\documentclass[12pt]{article}
\pdfoutput=1
\usepackage[pdftex,pagebackref,letterpaper=true,colorlinks=true,
            pdfpagemode=none,urlcolor=blue,linkcolor=blue,
            citecolor=blue,pdfstartview=FitH]  {hyperref}

\usepackage{amsmath,amsfonts}
\usepackage{graphicx}
\usepackage{color}

\newcommand{\half}{\textstyle{\frac{1}{2}}}

\DeclareMathOperator{\sgn}{sgn}

\begin{document}

\begin{center}
\begin{LARGE}
{\bf The Ping Pong Pendulum%
\footnote{243-PingPongPendulum.tex}%
}
\end{LARGE}

\begin{large}
{\bf Peter Lynch, UCD, Dublin}
\end{large}
\end{center}

\begin{narrower}
{\footnotesize {\bf Abstract:} 
Many damped mechanical systems oscillate with increasing frequency as the amplitude decreases.
One popular example is Euler's Disk, where the point of contact rotates with increasing rapidity 
as the energy is dissipated. We study a simple mechanical pendulum that exhibits this behaviour. }
\end{narrower}

Galileo noticed the regular swinging of a candelabra in the cathedral in Pisa and
speculated that the swing period was constant. This led him to use a pendulum to measure intervals
of time for his experiments in dynamics \cite[pg 97]{TOB03}.
Galileo's conclusion was close to correct: for small swing angles, the period of a pendulum varies little
with amplitude. Even with a swing through $180^\circ$, the period is only $18\%$ greater than that for a
small swing angle.

\section*{V-shaped Potential}

\begin{figure}[h]
\begin{center}
\includegraphics[scale=0.50]{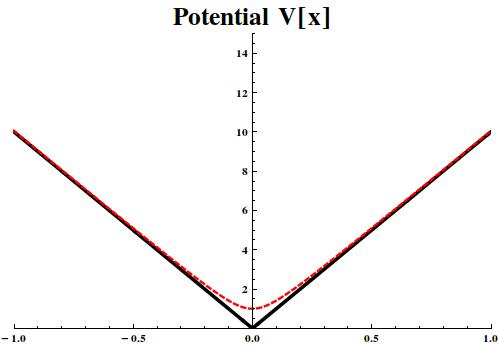}
\caption{V-shaped potential function $V(x) = \kappa |x|$ (solid black line) and a
regular approximation $V_\epsilon(x) = \kappa\sqrt{x^2+\epsilon^2}$ (dashed red line).}
\label{fig:Vpot}
\end{center}
\end{figure}

Unlike the simple pendulum, many mechanical systems rock back and forth with decreasing period
as the motion dies down, oscillating with ever-increasing frequency.
To exemplify this behaviour, let us look at a particle moving in a V-shaped potential well.
To be specific, let us take the potential energy to be
$$
V(x) = \kappa |x| \,.
$$
which is shown in Fig.~\ref{fig:Vpot}. Also shown is the function
$V_\epsilon(x) = \kappa\sqrt{x^2+\epsilon^2}$ which is regular and which approximates $V(x)$
as $\epsilon\rightarrow 0$.

For the V-potential well, the restoring force is
$$
F(x) = -\frac{\partial V}{\partial x} = -\kappa\sgn x = 
\begin{cases} -\kappa & \mbox{if } x > 0 \\
              +\kappa & \mbox{if } x < 0 \end{cases}
$$
The equation of motion for a particle of unit mass is $\ddot x = F(x)$
\cite{SandG59}. Since the force is constant as
long as the sign of $x$ remains unchanged, we can write the solution for initial
conditions $x(0)=x_0 > 0, \dot x(0)=v_0$ by solving $\ddot x = -\kappa$:
$$
x(t) = x_0 + v_0 t - \half \kappa t^2 \,.
$$
Thus, $x$ is a parabolic function of time.
This remains valid until $t=t_0=[v_0+\sqrt{v_0^2+2\kappa x_0}]/\kappa$.
We will assume that $v_0=0$. Then $x$ reaches zero at time $t_0=\sqrt{2x_0/\kappa}$.

The solution is continued beyond $t_0$ by solving $\ddot x = +\kappa$. This gives another
parabolic segment, inverted with respect to the first one and joined smoothly to it.
Note that $x$ and $\dot x$ are continuous at $t_0$ but the acceleration $\ddot x$ is
discontinuous there.

\begin{figure}[h]
\begin{center}
\includegraphics[scale=0.50]{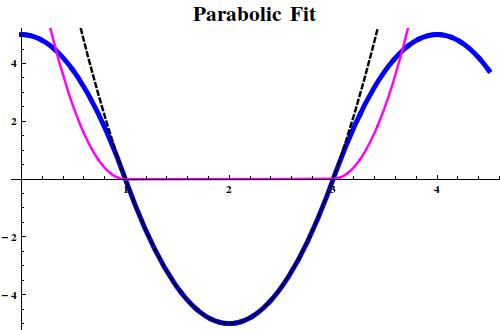}
\caption{Numerical solution with $x_0=5$ and $\kappa=10$ (heavy blue line),
and parabola fitting a segment (dashed black line).
Difference between the solution and the parabola (thin magenta line).}
\label{fig:parafit}
\end{center}
\end{figure}

The solution comprises a sequence of parabolic arcs smoothly joined at times
$\{t_0, 3t_0, 5t_0, \dots \}$. Fig.~\ref{fig:parafit} shows the numerical solution (heavy
blue curve) and a parabola fitted to the segment $(t_0,3t_0)$.  The thin magenta curve shows the
difference. One can see that the fit is excellent.

The motion is periodic, with period $\tau = 4t_0$ and frequency $\omega = 2\pi/\tau$, so
\begin{equation}
\tau = 4\sqrt{\frac{2 x_0}{\kappa}} \qquad\mbox{and}\qquad
\omega = \frac{\pi}{4} \sqrt{\frac{2\kappa}{x_0}} \,.
\label{eq:perfreq}
\end{equation}
We see immediately that $\omega\rightarrow\infty$ as $x_0\rightarrow 0$.

\section*{The Ping Pong Pendulum}

\begin{figure}[h]
\begin{center}
\includegraphics[scale=0.50]{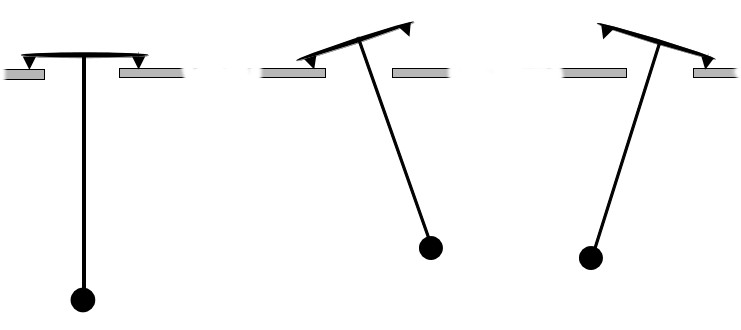}
\caption{The ping pong pendulum. Left panel: at equilibrium.
Centre panel: swinging to the right. Right panel: swinging to the left.}
\label{fig:PPP}
\end{center}
\end{figure}

\begin{figure}[h]
\begin{center}
\includegraphics[scale=0.60]{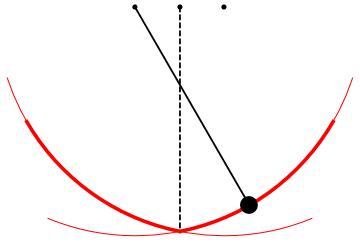}
\caption{Bob moves on a circular arc centered at the left pivot when swinging to the
right, and on an arc centered at the right pivot when swinging to the left.}
\label{fig:PPParcs}
\end{center}
\end{figure}

We consider a pendulum with two pivots (Fig.~\ref{fig:PPP}). The motion is confined to a
plane and the pendulum pivots alternately about each pivot.
The pendulum bob moves on a circular arc centered at the left pivot when it is 
swinging to the right (Fig.~\ref{fig:PPParcs}). When swinging to the left,
it rotates about the right pivot, following another circular arc. 
Thus, the motion of the bob is
along two circular arcs, intersecting at the point of equilibrium.

\begin{figure}[h]
\begin{center}
\includegraphics[scale=0.50]{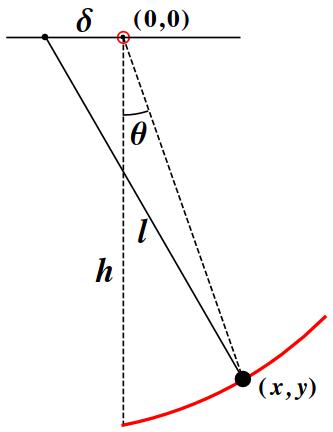}
\caption{Notation used to describe the system. The configuration is determined by the
angle $\theta$. The case $\theta>0$ is shown here.}
\label{fig:PPPsymbols}
\end{center}
\end{figure}

The origin of coordinates is taken to be at the point mid-way between the two pivot
points. The distance between the pivots is $2\delta$, the distance from pivot to bob is
$\ell$ and the point of equilibrium is at $(0,-h)$ (see Fig.~\ref{fig:PPPsymbols}).
The configuration of the system is determined by $\theta$, the angle
between the vertical through the origin and the line from origin to bob.
Elementary algebraic geometry shows that, for $\theta>0$, the coordinates of the bob are
\begin{eqnarray}
x &=& \phantom{-}\sqrt{\ell^2-\delta^2\cos^2\theta}\cdot\sin\theta - \delta\sin^2\theta
\label{eq:Xexp} \\
y &=&          - \sqrt{\ell^2-\delta^2\cos^2\theta}\cdot\cos\theta + \delta\sin\theta\cos\theta 
\label{eq:Yexp} 
\end{eqnarray}

\section*{Damping and Frequency Growth}

For small swing amplitudes, we may approximate the two circular arcs by line segments,
so that (\ref{eq:Xexp})--(\ref{eq:Yexp}) become 
\begin{equation}
x = h \theta \,, \qquad y = - h + \delta\theta \,.
\nonumber
\end{equation}
The tangents to the circular arcs at the point $(0,-h)$ are the lines $y=-h+(\delta/h)x$
and $y=-h-(\delta/h)x$.

The potential energy becomes a V-shaped potential well, $V(\theta) = \kappa|\theta|$,
where $\kappa=\delta/h$.
We have seen that the motion in such a potential well is described by a sequence of
parabolic arcs and the period is $\tau = 4\sqrt{{2 x_0}/{\kappa}} = 4\sqrt{{2 h \theta_0}/{\kappa}}$.

\begin{figure}[h]
\begin{center}
\includegraphics[scale=0.50]{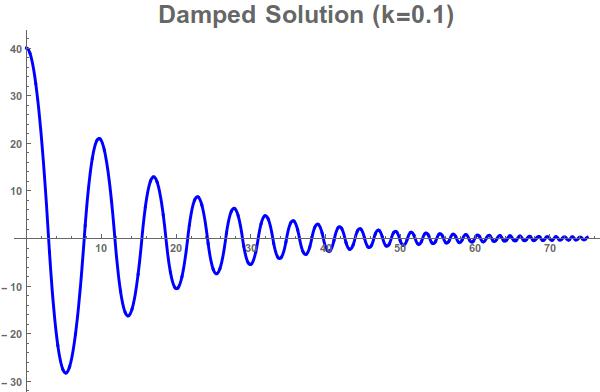}
\includegraphics[scale=0.50]{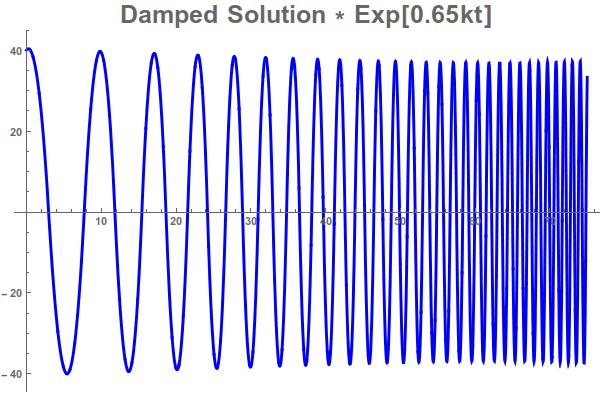}
\caption{Top: Solution of equation (\ref{eq:dampeqn}) with damping coefficient $k=0.1$.
         Bottom: Solution of equation (\ref{eq:dampeqn}) scaled up by an exponential factor.}
\label{fig:DampedSoln}
\end{center}
\end{figure}

Now we add damping to the system, and model the motion by the equation
\begin{equation}
\ddot\theta + k\dot\theta + \kappa \sgn\theta = 0
\label{eq:dampeqn}
\end{equation}
This equation can be solved piecewise in analytical terms. For $\theta>0$ the solution is
$$
\theta = \theta_0 + \left(\frac{\dot\theta_0}{k}+\frac{\kappa}{k^2}\right)(1-e^{-kt}) - \frac{\kappa t}{k} \,.
$$
However, it is inconvenient to have separate solutions for separate segments, so
we solve the equation numerically, replacing $V(x)$ by $V_\epsilon(x)$ with a small value of $\epsilon$.
The result shown in Fig.~\ref{fig:DampedSoln}.
It is clear that the frequency increases strongly as the amplitude decreases.
To make this even clearer, we also show 
the solution scaled by a growing exponential function.

This pattern of frequency increasing as energy decreases is
similar to the behaviour of a range of physical systems \cite{BandB05}.
We have mentioned the Euler Disk, but there are many others.

\bigskip\phantom{.}

\end{document}